\documentstyle[aps,prl,multicol]{revtex}
\input{psfig}

\newcommand{\bea}{\begin{eqnarray}}
\newcommand{\eea}{\end{eqnarray}}
\renewcommand{\[}{\begin{equation}} 
\renewcommand{\]}{\end{equation}}
\newcommand{\bef}{\begin{figure}} 
\newcommand{\ef}{\end{figure}}

\newcommand{\eg}{{\it e.g.}}
\newcommand{\llabel}[1]{\label{#1}}
\newcommand{\eq}[1]{Eq.~(\ref{#1})} 
\newcommand{\fig}[1]{Fig.~\ref{#1}}

\begin{document}
\maketitle

\vspace{-2cm}
\noindent
{\small Submitted to Special Issue of Superlattices and
Microstructures:\\ \underline{Third NASA Workshop on Device 
Modeling, August 1999.}}

\vspace{0.8cm}
\begin{center}
{\large \bf Shrinking limits of silicon MOSFET's: Numerical study of
10-nm-scale devices} 

\vspace{0.4cm}
{\large Y. Naveh\footnotemark\footnotetext{e-mail: yehuda@hana.physics.sunysb.edu} and K. K. Likharev} \\
{\it Department of Physics and
Astronomy, State University of New York, Stony Brook, NY
11794-3800}\\
(\today)
\end{center}


\renewcommand{\[}{\begin{equation}}
\renewcommand{\]}{\end{equation}}

\begin{abstract}
We have performed numerical modeling of dual-gate ballistic
$n$-MOSFET's with channel length of the order of 10 nm, including the
effects of quantum tunneling along the channel and through the gate
oxide. Our analysis includes a self-consistent solution of the full
(two-dimensional) electrostatic problem, with account of electric
field penetration into the heavily-doped electrodes.
The results show that transistors with channel length as small
as 8 nm can exhibit either a transconductance up to 4,000 mS/mm or
gate modulation of current by more than 8 orders of magnitude,
depending on the gate oxide thickness. These characteristics make the
devices satisfactory for logic and memory applications, respectively,
though their gate threshold voltage is rather sensitive to
nanometer-scale variations in the channel length.
\end{abstract}

\begin{multicols}{2}

\section{Introduction}

For almost half a century, the number of silicon MOSFET's on
a single commercial chip has approximately doubled every eighteen
months. There is little doubt that this exponential growth will
continue throughout the next decade, allowing for a
minimum linear scale of 50 nm at around 2010\cite{roadmap}. At shorter lengths,
significant 
technological problems arise. However, silicon MOSFET's with channel
length approaching 30 nm have been studied widely in the past few
years, both theoretically and experimentally (for a review see
Refs.~\onlinecite{Taur 97,Wong 99}).

A further reduction of MOSFET's linear size in bulk production will
inevitably require radical advances in lithography and doping
technologies, and possibly also require a change in the basic
transistor geometry. However, 
we are not aware of any fundamental
limitation on the performance of sub-30 nm scale MOSFET's. In fact, it
was very recently shown\cite{Huang 99} that a double-gate silicon-on-oxide (SOI)
transistor with a gate length of 18 nm can exhibit an acceptable
$I_{on}/I_{off}$ current ratio of about eight orders of magnitude. In
the present work we study theoretically transistors with even shorter
gates, of length between 3 and 10 nm. We show that such devices (with
channel length as short as 8 nm) can exhibit performance sufficient for
both logic and memory applications.

Three fundamental differences exist between ordinary, 100-nm-scale
MOSFET's, and short, 10-nm-scale devices. First, as the short channel
transistors are of length comparable to or smaller than the scattering
mean free path of electrons in the channel, electron transport in
them is essentially ballistic. This is in contrast to drift-diffusion transport in
long transistors. Second, at 10-nm length scales, quantum mechanical
tunneling can be significant, and as will be shown below, may actually
dominate at some parameter range. Lastly, as the short
transistors are of length comparable to the electrostatic screening
length of electrons, one dimensional approximations are not sufficient
to describe the electrostatic potential in the system, and a full,
two-dimensional solution of the Poisson equation is
required. Moreover, in contrast to long devices, we will show that
finite penetration of electric field into
the source, drain, and gate electrodes can crucially affect 
the source-drain current in the 
short-channel MOSFET's, and cannot be neglected.

Previous works attempting to model short-channel MOSFET's usually
took into account the ballistic nature of electron transport, but relied on
one-dimensional approximations, and neglected electron tunneling and
finite fields in the electrodes. Natori\cite{Natori 94} was
the first to employ this approach. He calculated the I-V
characteristics of both single-gate and dual-gate MOSFET's, and 
showed that, within this simplistic approach, the current exhibits a strong
saturation due to the exhaustion of all source electrons by large
source-drain voltage. Within the same
approximations, Lundstrom\cite{Lundstrom 97} arrived at a 
phenomenological equation which allows for a treatment of finite
backscattering of electrons. This 
approach was extended very recently\cite{Assad 99}, and the
on-current, and in particular the effects of higher subbands on the
current, were studied. However, short-channel effects such as
drain-induced barrier lowering (DIBL) cannot be described within this
one-dimensional approximation, and therefore the authors of
Ref.~\onlinecite{Assad 99} limited themselves to devices of the order
of 100 nm. A previous study by Pikus and Likharev of short-channel ballistic
MOSFET's\cite{Pikus 97} also relied on a one-dimensional approximation
of the electrostatics in the system, and neglected quantum tunneling
and backscattering. Thus, the results of that work (mainly, that a 5-nm
gate-length transistor may perform sufficiently well) could be
questioned. Finally, Wong {\it et.\ al.\ }\cite{Wong 98} 
studied ballistic MOSFET's with a somewhat longer minimal channel
length of around 15
nm. They concentrated on two-dimensional short channel effects,
but studied only the closed state (so only the Laplace equation had to be
solved). At these length scales neglect of
tunneling is justified.

The aim of the present work was to model the transport in ultrashort
silicon MOSFET's, with channel length of around 10
nanometers. Our modeling uses as few simplifying assumptions as
possible, the main one being the neglect of backscattering, an
assumption which is justified by existing mobility data, and by our
choice of transistor geometry -- see the next section. We take full
account of the ballistic nature of transport in the transistor and of
quantum mechanical tunneling. We solve the full, two-dimensional,
Poisson equation not only in the channel, but also in all
electrodes, thus allowing for charging effects inside the source, drain,
and gates. We study both the {\it on} and {\it off} states.

\section{Model}

\begin{figure}
\centerline{\hspace{8cm} \psfig{figure=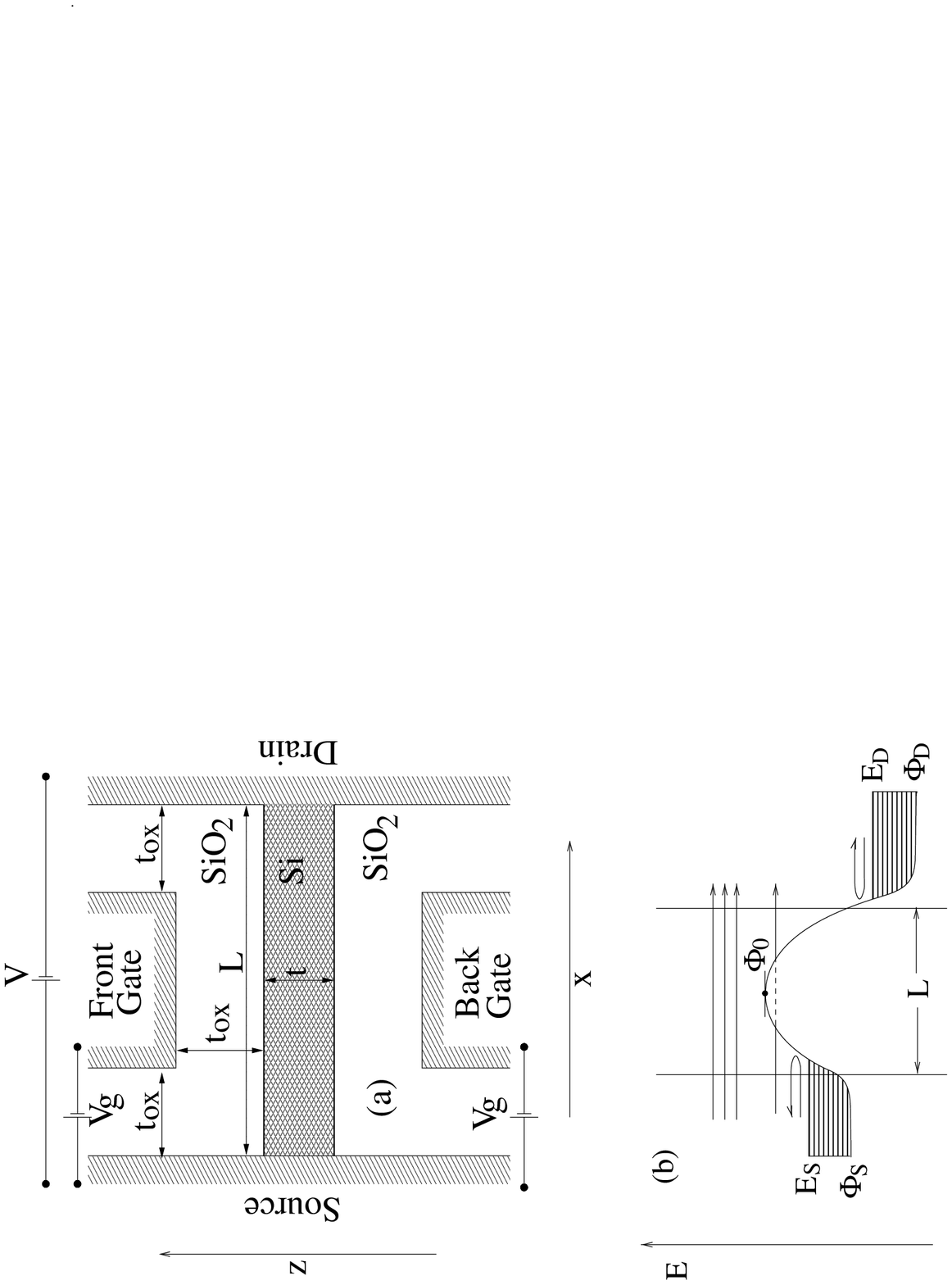,angle=-90,width=150mm}}
\caption{\label{f:model} \narrowtext (a) The model of dual-gate MOSFET used in
this work. (b) 
General scheme of transport in ballistic transistors.}
\end{figure}

The geometry of the transistor under consideration in this paper is
shown schematically in \fig{f:model}(a). As appears to be the consensus on
the optimal general design for ultrasmall MOSFET's\cite{Taur 97,Wong
99,Huang 99}, we study here a double-gate SOI structure. A thin silicon
channel of length $L$, width $W$, and thickness $t \ll W$ 
connects two bulk source and drain polysilicon electrodes. The
electrodes are heavily doped to a donor density $N_D$, while the
silicon layer is undoped. Both the electrodes and the
channel are separated from two polysilicon gate electrodes, doped to a
density $N_G$, by silicon oxide layers of identical thickness
$t_{ox}$. Source-drain voltage $V = E_D - E_S$ and gate voltage $V_g =
E_G - E_S$ are determined from the Fermi levels $E_S$, $E_D$, and
$E_G$ in the source, drain, and gate electrodes, respectively. A typical
energy profile $\Phi(x)$ in the center of the silicon layer is shown
in \fig{f:model}(b). $\Phi_0$ denotes the peak of the
potential energy in the channel.

Motion in the thin silicon channel is affected by lateral quantum
confinement. In this work we consider only the case where the channel 
is so thin that $E_S$, $E_D$, and temperature $T$ are well
below the second subband energy, so only the first subband
participates in transport. This assumption leads to an important
aspect of our model: due to the mismatch in phase space between the
strictly 2D channel and the bulk, 3D source and drain electrodes, an
electron impinging from the channel on one of the electrodes would be
absorbed by that electrode, {\it i.e.}, 
would have only a negligible probability of backscattering (from
source/drain 
impurities or phonons) into the
channel\cite{Solomon 99}. If the scattering mean free path in the
electrodes is larger than $t$ then, due to the mismatch also in real space, the
already negligible backscattering
probability becomes even smaller.

Thus, in our geometry, the usual transport model of ballistic
transistors\cite{Natori 94} holds to a very good approximation. Within
this model electron distribution inside the source and drain
electrodes is the equilibrium distribution at the lattice temperature
$T$. Source electrons impinging on the channel are absorbed by it,
and then travel ballistically 
inside the channel. However, only electrons with energy higher than
$\Phi_0$ are
certain to 
arrive at (and be absorbed by) the drain electrode\cite{Pikus
97}. Electrons with
lower energy are transmitted through the potential 
barrier
with the quantum 
mechanical tunneling probability $\Theta < 1$, and 
reflected from it with probability $1 - \Theta$.  The reflected
electrons  travel back ballistically towards the source electrode,
where they are absorbed [see \fig{f:model}(b)]. A similar (but opposite) process
describes the drain electrons. 

All electrons in the channel and in the electrodes, as well as the
ionized donors in the electrodes, contribute to the local
electrostatic potential. The self consistent solution of this
Poisson-transport problem is at the heart of our
calculations\cite{neglecttunnel}. Once the local potential and
electron density are known for any given set of $V$ and $V_g$, the
source-drain current $I$ can be calculated as the difference between
the 
source-to-drain current and the drain-to-source current\cite{Natori 94,Lundstrom
97,Pikus 97}. The gate current due to oxide leakage can also be
calculated within the regular tunneling approach.

\section{Theory}

The formulation of the theory describing the above model is similar to
previous works by Natori\cite{Natori 94}, 
Lundstrom and co-workers\cite{Lundstrom 97,Assad 99}, and especially by
Pikus and Likharev\cite{Pikus 97}. The two main differences in the
theory between
this and the previous works
are that
there is no attempt here to reduce the full Poisson equation to a
one-dimensional approximation, and the source-drain tunneling current
is now taken into account. Also, in the present work the quantization
energy $E_1$ needs to be taken explicitly into account because it is
present only in the channel.

The basic equation to be solved in the channel, the oxide layers, and
the source, drain, and gate electrodes is the Poisson equation for
$\Phi(x, z)$, where $x$, $z$ are the directions along and 
perpendicular to the channel length, respectively (see \fig{f:model}):
\[ 	\label{Poisson}
\nabla^2 \Phi(x, z) = {4 \pi \over \epsilon(x, z)} \rho(x, z).
\]
Here $\epsilon(x, z)$ is the local dielectric constant and $\rho(x,
z)$ the local three dimensional charge density. The boundary conditions
for \eq{Poisson} at $x,z \to \pm \infty$ are given in terms of the applied
voltages $V$, $V_g$ which determine the difference between the  
Fermi levels $E_S$, $E_D$, and $E_G$ (The boundary conditions are
taken at distances much larger than the screening length in the
electrodes, where charge neutrality allows for a unique determination
of the 
local potential from the known values of the local Fermi energies and donor
densities). 

The charge density in the structure is given by
\end{multicols}
\widetext
\begin{eqnarray}	\label{density}
 \rho(x, z)  =  
& &  \left\{ \begin{array}{ll}
-{2 q \over t} \cos^2 \left( {\pi \over t} z \right) n_2 \left[
\Phi(x,z) \right]  
& \left[ (x, z) \in {\cal C} \right] \\
q N_{D,G} - {3 \left(2 \bar m \right)^{3/2} q\over \pi^2 \hbar^3}
\int_{\Phi(x,z)}^\infty {\left[ E - \Phi(x, z) \right]^{1/2} \over \exp \left( {E - E_{F}
\over T} \right) + 1} \, dE
 & \left[ (x,z) \in {\cal E} \right] \\
0 & {\rm otherwise,} 
\end{array} \right.
\end{eqnarray}
\begin{multicols}{2}
Where ${\cal C}$, ${\cal E}$ denote
the channel and electrode (source, drain, and gate) regions,
respectively, $E_F$ is the Fermi level in the corresponding electrode, 
$\bar{m}$ is the three-dimensional density-of-states effective mass,
and $q=|q|$ is the electron charge. The
two-dimensional electron density $n_2$ is the sum of all partial
densities of electrons at energy $E_x = \Phi(x,z) + m_1 [v_x(x,z)]^2 / 2$, with
${\bf v}$ the electron velocity and $m_1$ the mass of the electron in the direction of the
current [which, due to quantization (and assuming $100$ orientation), is
the light electron mass]. Counting the transmitted and reflected
electrons, both from the source and the
drain electrons, $n_2$ is given by\cite{Natori 94,Pikus 97}:
\[ \llabel{n2}
n_2^{\pm}\left[E_x, \Phi(x) \right] = -{J^\pm(E_x) \over q v\left[E_x, \Phi(x)
\right]},
\]
with $J^\pm(E_x)$ the partial current density in each direction. $J^\pm(E_x)$
is uniform throughout the length of the structure and can be
calculated by using the equilibrium distribution in the source and
drain electrodes.
In \eq{density} $n_2$
is multiplied by the cosine quantum factor in order to obtain the 3D
density in the channel. In the electrodes, \eq{density} uses the
equilibrium Fermi-Dirac density of electrons, while in the oxide
$\rho$ is taken to be zero.

Once equations (\ref{Poisson}) and (\ref{density}) are solved
simultaneously, the source-drain current can be calculated as the sum
over all partial currents, each multiplied by the energy-dependent
quantum transmission probability through the potential in the channel:
\[ \llabel{current}
J = {q \over \pi^2 \hbar} \int_{-\infty}^\infty dk_y \, \int_0^\infty dE_x \, \Theta (E_x)
\left[ f_0(E) - f_0(E + V) \right],
\]
where $f_0$ is the Fermi function, 
\[  \llabel{totalE}
E = E_1 + E_x + {\hbar^2 k_y^2 \over 2 m_1},
\]
and the transmission probability is given by the WKB result
\end{multicols}
\widetext
\begin{eqnarray}	\label{transmission}
 \Theta (E_x)  =  
& &  \left\{ \begin{array}{ll}
\exp \left\{ -2 \int_{x_1}^{x_2} {\sqrt{2m_1 \left[\Phi(x) + E_1 -
E_x \right]} \over \hbar} \, dx \right\}
& E_x < \Phi_0 + E_1, \\
1 & {\rm otherwise,} 
\end{array} \right.
\end{eqnarray}
\begin{multicols}{2}
with $x_{1,2}$ the classical turning points.

We solve Eqs.~(\ref{Poisson}) and (\ref{density}) using a Poisson
solver designed specifically for this problem. As we will see, for
short devices screening in the electrodes may have a strong effect on
the performance of the device. Our solver therefore treats the
electrodes (source, drain, and gates) on an equal footing with the
channel. This requires that we use a very small mesh size, of the
order of 0.1 nm. Nonetheless,  our program
(which uses the conjugate
gradient method\cite{Press 92} as its basic algorithm) works rather
efficiently even with such a dense mesh. For 10-nm-scale devices, it takes
a low-end workstation CPU run time of only about 3 seconds for a
single iteration and less than 1 minute for the full calculation at a
single parameter set.

\begin{figure}
\centerline{\hspace{9.7cm}
\psfig{figure=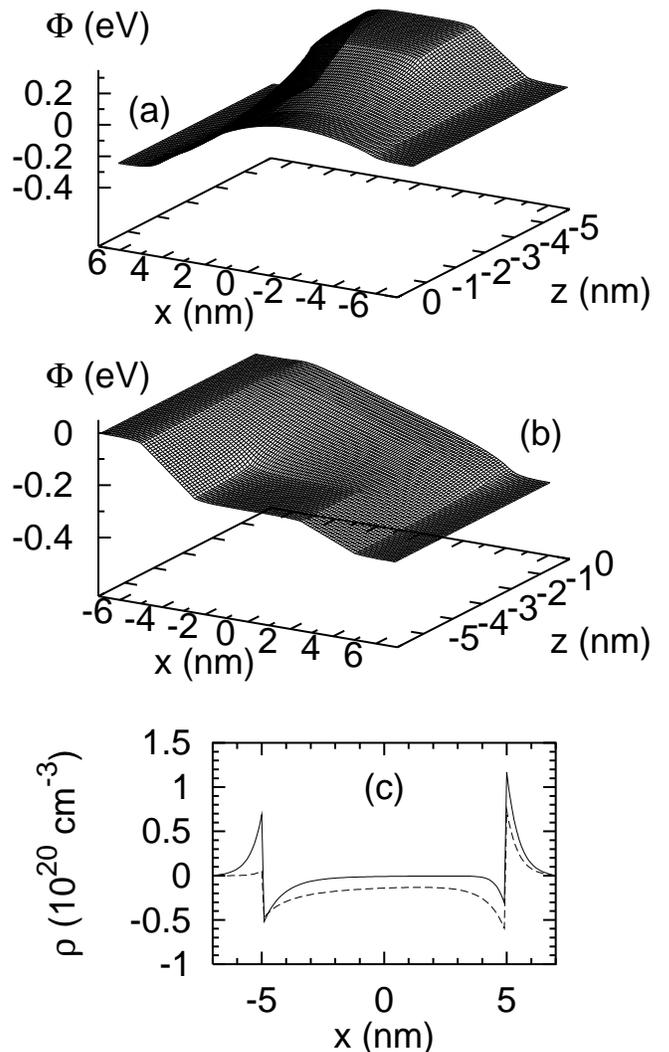,angle=-90,width=190mm}}
\caption{\label{f:potential} \narrowtext Electric potential distribution in the lower
half of the 
transistor for (a) typical 
negative gate voltage ($V_g = -0.3$ V) and (b) typical 
positive gate voltage ($V_g = 0.2$ V). (c) Charge density
distributions in the center of the channel for $V_g = -0.3$ V (solid
line) and $V_g = 0.2$ V 
(dashed line). In all three panels $L = 10$ nm,
$t_{ox} = 2.5$ nm, and $V = 0.3$ V.}
\end{figure}

Examples of the results of these calculations are shown in \fig{f:potential},
where the potential energy $\Phi$ is plotted as a function of $x$ and
$z$ for typical cases of {\it off} and {\it on} states at a finite
source-drain voltage. [only $z < 0$ is shown, as $\Phi(x,z) = \Phi(x,
-z)$ due to the symmetry of our geometry]. The two-dimensional short
channel effects (here $L = 10$ nm) clearly manifest themselves in
this figure: first, it is seen that in the {\it off} state [\fig{f:potential}(a)]
$\Phi_0$ assumes a value significantly smaller than $-q V_g$. Second, in
the {\it on} state [\fig{f:potential}(b)] $\Phi_0$ vanishes, and there is no longer a
potential barrier in the channel. The latter effect accounts for the
saturation of the current described in Refs.~\onlinecite{Natori
94,Lundstrom 97}. \fig{f:potential}(c) shows the charge distribution
at $z=0$ for the same gate voltages as in panels (a,b).

\section{Results}

In this paper we present results for silicon $n$-MOSFET's with what
we consider to be the optimal donor density and channel thickness (see
the next section for a discussion) of $N_D = 3 \times 10^{20}$ cm$^{-3}$
and $t = 2$ nm. For these parameters, $E_1 = 96$ meV, while $E_F
= 150$ meV. We study two different oxide thicknesses: the
'thin-oxide' device ($t_{ox}=1.5$ nm) will be shown to be suitable for logic
applications (for which small gate leakage is tolerable), while the
'thick-oxide' transistor ($t_{ox} = 2.5$ nm) will be suitable for
memory applications, in which it is necessary to have many orders of
magnitude control of the subthreshold current, while voltage gain is
of minor importance. We present all results for both  'short' ($L =
8$ nm) and 'long' ($L = 12$ nm) devices.

\begin{figure}
\centerline{\hspace{6.3cm} \psfig{figure=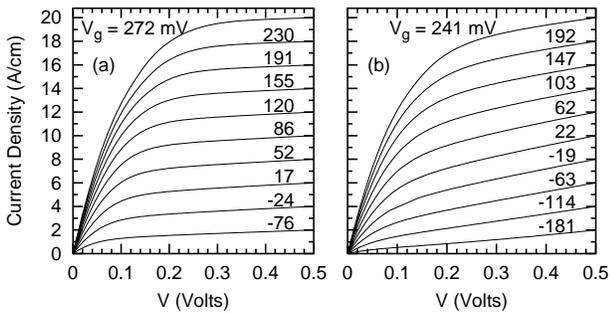,angle=-90,width=150mm}}
\caption{\label{f:J-Vthin}Source-drain $I-V$ curves of transistors with (a) $L = 12$ nm
and (b) $L= 8$ nm, for 10 values of gate voltage. Oxide thickness
is $t_{ox} = 1.5$ nm. }
\end{figure}

Figure \ref{f:J-Vthin} shows
$I-V$ characteristics of the 'thin-oxide' device 
for ten different values of gate voltage. The curves show a well-expressed
current saturation even at $L = 8$ nm, while at $L=12$ nm the
saturation is almost flat.  In these ultrasmall devices the saturation
shows up only when the electron potential energy maximum in the
channel is suppressed by positive gate voltage, and is due to the
exhaustion of source electrons \cite{Natori 94,Lundstrom 97}. However,
an effect which wasn't taken into account in the earlier works  
\vspace{1cm}
\noindent
is the
finite screening in the source and drain electrodes. As the screening
length in the electrodes becomes comparable to the channel length, the
voltage no longer falls only on the channel. This leads to a
significant decrease of the potential energy at the source-channel interface
with increasing source-drain voltage. Thus, the quantization energy
$E_1$ (relative to the source Fermi energy $E_S$) is lowered with increasing
$V$, an effect which accounts for the increase of current with $V$
(\fig{f:J-Vthin}) even in the 'totally saturated' regime.

\begin{figure}
\centerline{\hspace{6.3cm} \psfig{figure=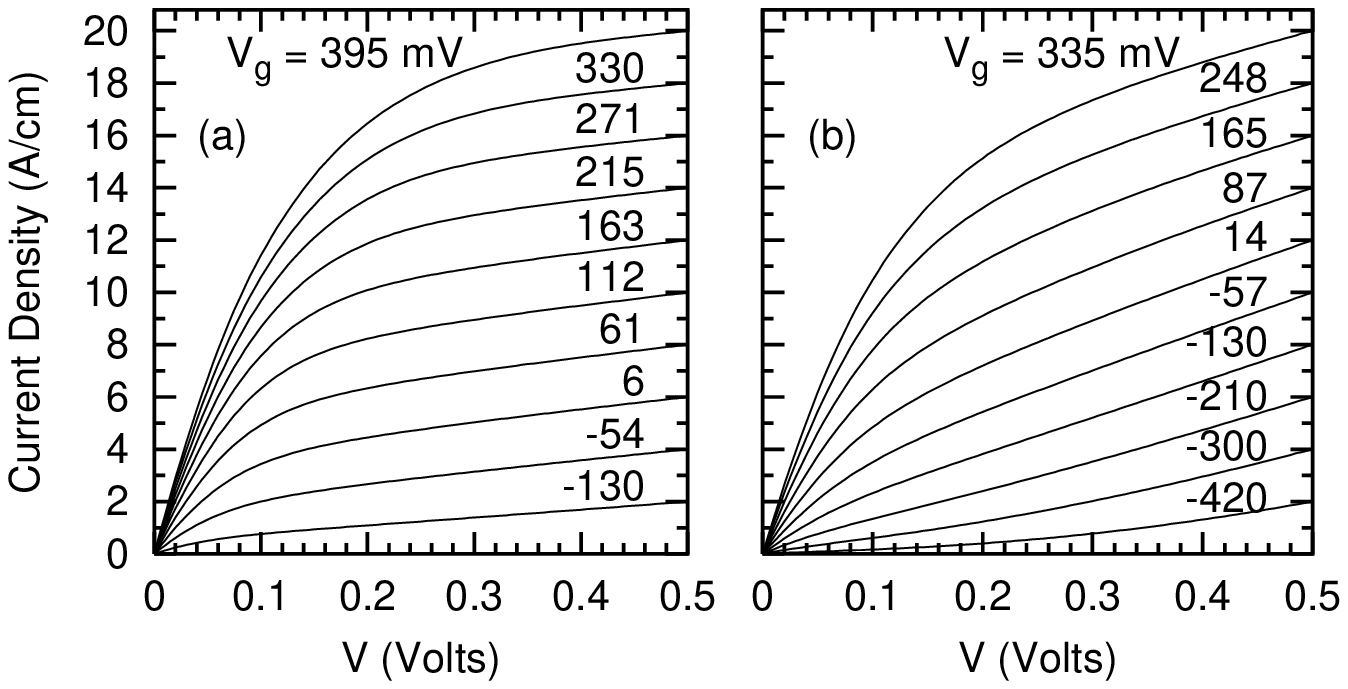,angle=-90,width=150mm}}
\caption{\label{f:J-Vthick}Same as in \fig{f:J-Vthin}, but for
$t_{ox} = 2.5$ nm.}
\end{figure}

Figure \ref{f:J-Vthick} shows the same characteristics as \fig{f:J-Vthin}, but for the
'thick-oxide' transistor. The characteristics now are much more
linear, and saturation  practically vanishes in the 8-nm
length device.

\vspace{-6cm}

\vspace{-6cm}

\begin{figure}
\centerline{\hspace{4.5cm} \psfig{figure=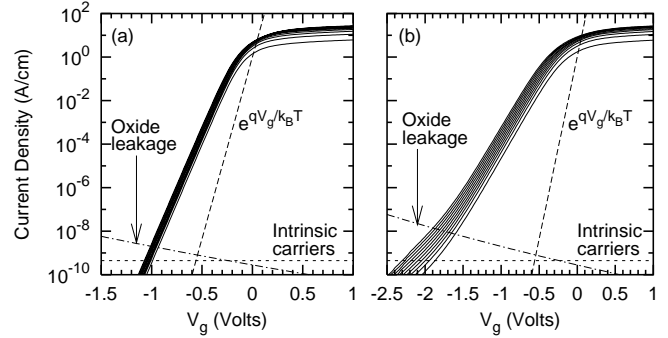,angle=-90,width=150mm}}
\caption{\label{f:subthreshthick}Subthreshold curves for transistors with (a) $L = 12$ nm
and (b) $L= 8$ nm for 10 values of source-drain voltage between $V =
0.03$ V and $V = 0.3$ V. 
Almost-vertical dashed lines denote the 60-mV-per-decade slope of an
ideal transistor. 
Horizontal dashed
lines represent the 
limit below which
intrinsic carriers are not negligible. Dot-dashed lines show the
current due to tunneling through the gate oxide.}
\end{figure}

\begin{figure}
\centerline{\hspace{4.5cm} \psfig{figure=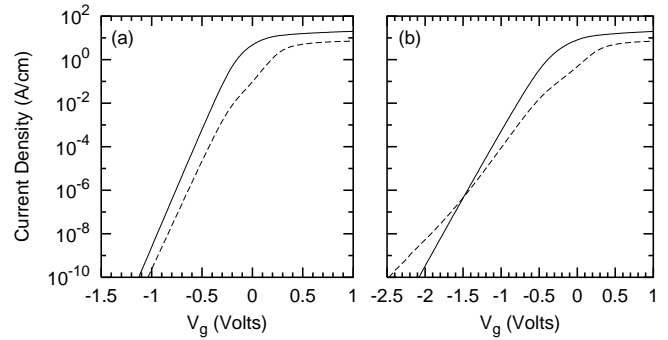,angle=-90,width=150mm}}
\caption{\label{f:separate}Thermal (solid lines) and tunneling
(dashed lines) currents as function of gate voltage at (a) $L = 12$ nm
and (b) $L= 8$ nm. Here $V = 0.3$ V. All other parameters are as in
\fig{f:subthreshthick}.}
\end{figure}

Sub-threshold curves of the 'thick-oxide' transistors are presented in
\fig{f:subthreshthick} for ten different source-drain voltages. For the 12-nm
device the curves have a nearly perfect log slope (indicated by the
dashed line) and very
small DIBL effect. However, the slope rapidly goes down and DIBL up
as the length decreases below ~10 nm. This loss is especially rapid at small
currents (big
negative gate voltages) due to electron tunneling under the narrower "bump"
in the electric
potential profile. Figure \ref{f:separate} shows the tunneling and thermal
currents for the same parameters as in \fig{f:subthreshthick} separately
(the sum of these two 
components gives the current presented by the upper curves of \fig{f:subthreshthick}). It is clear from the
figure that the tunneling current dominates the subthreshold current
at small $L$ ($\sim 8$ nm) and large negative $V_g$. 
In fact, the tunneling clearly affects not only the magnitude of the
current in the {\it off} state, but also the qualitative shape of the
subthreshold curve, which is no longer exponential -- see
\fig{f:subthreshthick}(b). 

Also shown in \fig{f:subthreshthick} are the oxide-leakage current and the current due
to intrinsic carriers (both are evaluated within a simple model and
should be taken only as an order of magnitude estimate). Due to the
large effective gap implied by the quantum confinement of electrons
and holes, the latter is very small compared to the former, so
gate-oxide leakage becomes the main limiting mechanism on the
subthreshold performance. Another deteriorating effect, Zener
tunneling of holes from the drain electrode into the channel, appears only at
negative gate voltages much larger than the ones we consider here
(where the oxide leakage current is already prohibitive).

\vspace{-5.5cm}

\vspace{-6.0cm}

\begin{figure}
\centerline{\hspace{4.5cm} \psfig{figure=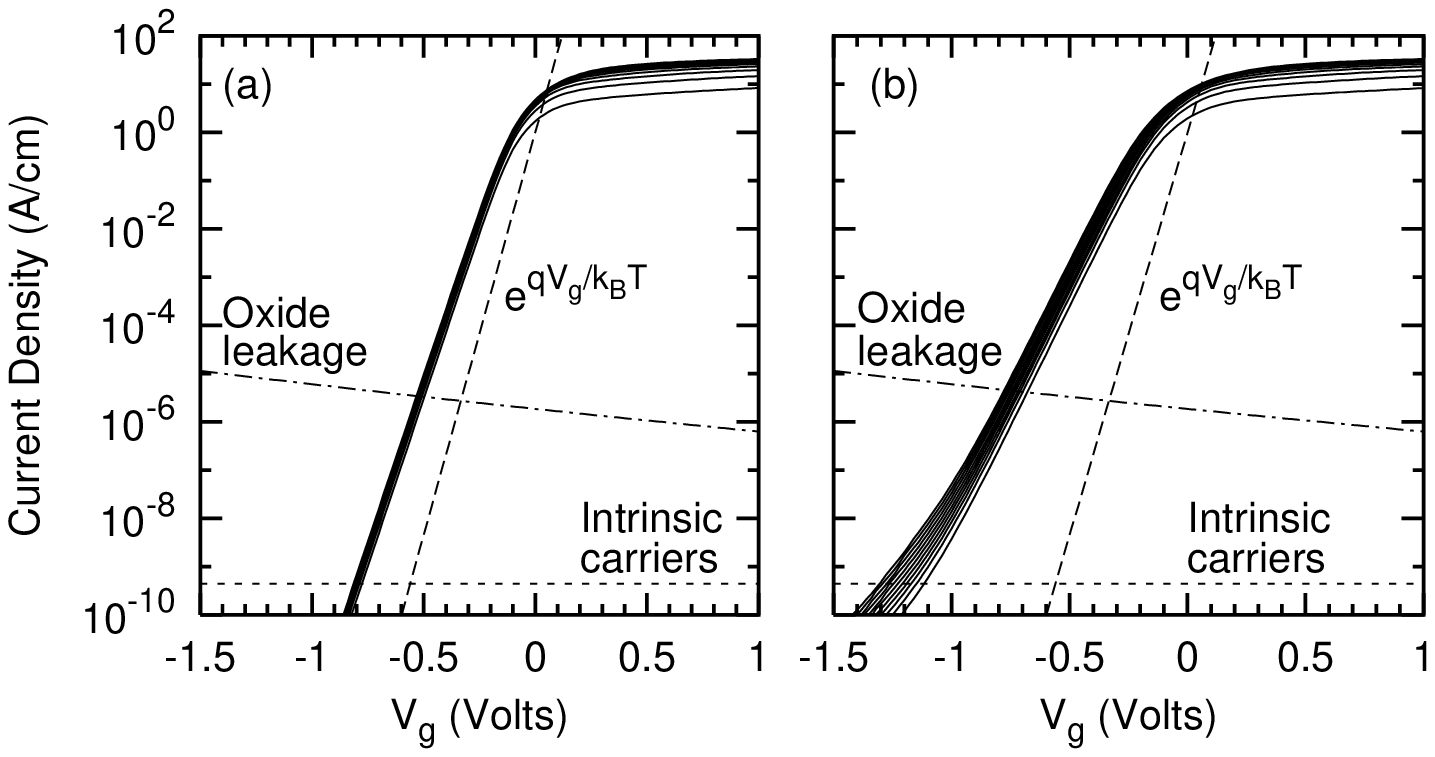,angle=-90,width=150mm}}
\caption{\label{f:subthreshthin}Same as in \fig{f:subthreshthick}, but
for $t_{ox} = 1.5$ nm.}
\end{figure}

Figure \ref{f:subthreshthin} shows subthreshold curves for the 'thin-oxide' device. The
slopes of the curves in this case are almost ideal even for the
short-channel device [\fig{f:subthreshthin}(b)]. However, the large oxide leakage
means that the $I_{on}/I_{off}$ ratio is reduced to $10^{-6}$.

\begin{figure}
\centerline{\hspace{11.5cm} \psfig{figure=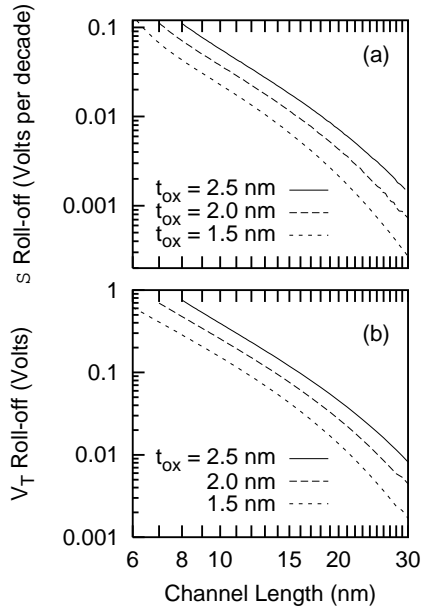,angle=-90,width=180mm}}
\caption{\label{f:rolloff}(a) Subthreshold slope roll-off and (b) threshold
voltage roll-off as a function of sample length for various oxide
thicknesses.}
\end{figure}

In addition to the $I_{on}/I_{off}$ ratio, there exist at least three other
figures of merit which characterize the subthreshold
curves. First is the subthreshold slope roll-off ${\cal S}-{\cal
S}_{id}$, with ${\cal
S}_{id} = 60$ mV/decade the ideal room-temperature slope. This
roll-off is plotted in \fig{f:rolloff}(a)
as a function of $L$, for three values of $t_{ox}$. 

Second is the 
threshold voltage roll-off $V_T-V_T^\infty$ where $V_T$ is defined here
as the gate voltage at which $J=2 \times 10^{-4}$ A/cm and $V_T^\infty$ is the
threshold voltage at $L \to \infty$. In our geometry, $V_T^\infty =
313$ mV. The $V_T$ roll-off is shown in \fig{f:rolloff}(b) as a function of $L$
for the same values of $t_{ox}$ as in \fig{f:rolloff}(a).

\vspace{-5.5cm}

\vspace{-4.5cm}

\begin{figure}[t]
\centerline{\hspace{11.5cm} \psfig{figure=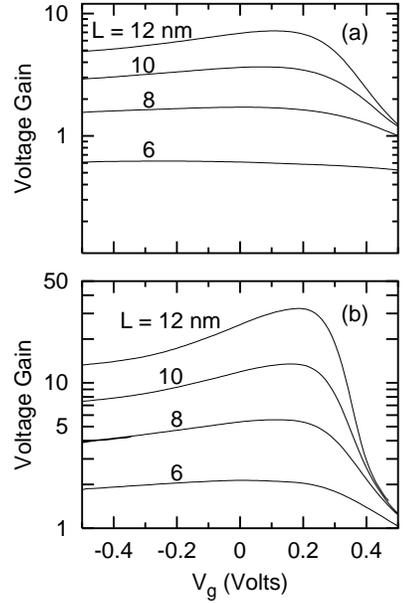,angle=-90,width=180mm}}
\caption{\label{f:gain}Voltage gain as a function of gate voltage for various
channel lengths. (a)  $t_{ox}=2.5$ nm. (b)  $t_{ox}=1.5$ nm.}
\end{figure}

Lastly, the finite value of voltage gain, defined as $G_V = \left(
dV/dV_g\right)_{I={\rm const}}$, (in contrast to the ``ideal'' value
of infinity) can be used to characterize 
both DIBL at the subthreshold state and imperfect saturation at the
open state. Voltage gain as a function of gate voltage for both the
'thin-oxide' and 'thick-oxide' devices, and for various $L$'s, is
shown in \fig{f:gain}. In order to evaluate the results here, one should
remember that the 
usual CMOS design tools
imply $G_V \gg 1$, while devices with $G_V < 1$ cannot sustain logic
circuits.

\begin{figure}
\centerline{\hspace{11.5cm} \psfig{figure=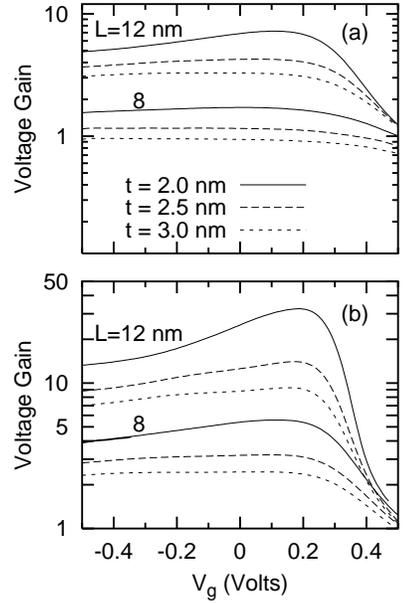,angle=-90,width=180mm}}
\caption{\label{f:thickt} Same as in \fig{f:gain}, but for various
channel thicknesses. (a)  $t_{ox}=2.5$ nm. (b)  $t_{ox}=1.5$ nm.}
\end{figure}

In order to show the effect of channel thickness on the results, we
reproduce in \fig{f:thickt} the plots of \fig{f:gain} at \\ {\tiny .}
\vspace{-4.9cm}

\vspace{-4.5cm}
\noindent
$L=8$ nm and $L=12$ nm, and
include also the cases of $t = 2.5$
nm and $t= 3$ nm. At $t=2.5$ nm our model is still strictly
valid. For the sake of understanding the
thickness-dependence of the performance of the device, we also present
results for $t = 3$ nm, at which $E_2 = 170$ meV is higher than
$E_S$ by only 20 meV. At this and larger values of $t$, the second
quantized level becomes too  
close to the source Fermi energy, so transport through the second
subband (possible at finite bias via tunneling of source electrons
into the second subband inside the channel) may not be negligible.

\section{Discussion and Conclusions}

The main conclusion which can be drawn from our results is that
ballistic, dual-gate
transistors with channels as short as 8 nm still seem suitable for
digital applications, with proper choice of the gate oxide thickness.
In fact, devices with a relatively thick oxide allow a very high
$I_{on}/I_{off}$ ratio, above 8 orders of magnitude [\fig{f:subthreshthick}(b)],
making them suitable for memory applications including both DRAM and
NOVORAM \cite{Likharev 99}. In contrast, transistors with thinner gate
oxides (say, 1.5 nm) have a gate oxide leakage too high for memory
applications [\fig{f:subthreshthin}(b)], but their transconductance of about 4000 mS/mm
[\fig{f:J-Vthin}(b)] and voltage gain of around 5 over a wide range of gate
voltages [\fig{f:gain}(b)] are sufficient for logic circuits. The
performance of the devices, both for logic and memory applications,
improves dramatically when going from $L = 8$ nm to $L = 12$ nm
[Figs.~\ref{f:J-Vthin}(a),\ref{f:subthreshthick}(a),\ref{f:gain}(b)].
The results
presented here are compatible with the results of
Ref.~\onlinecite{Pikus 97} (when comparing gate lengths),
Ref.~\cite{Wong 98} (in which MOSFET's with larger $t$ have been
studied), and Ref.~\onlinecite{Assad 99} (in which the open-state
current in longer devices has been calculated within a simplified 1D
model.)

At least in the geometry which we study here, a channel length of 8 nm
seems to be very close to the lower limit of still-feasible
MOSFET's. Most importantly, the maximal voltage gain drops to 2
already at 6 nm even for the 'thin-oxide' device [\fig{f:gain}(b)], a fact which
renders it useless for logic applications. As for memory applications,
they are basically limited by the minimal thickness of the gate
insulator. As long as the insulator used is SiO$_2$, $t_{ox}$ cannot be
significantly thinner
than 2.5 nm, which is the thinnest layer which still gives an
eight-orders-of-magnitude control over the current
[\fig{f:subthreshthick}(b)]. This implies a strict limit of 5 nm on the 
channel 
length. However, a working device of $L = 6$ nm is hard to
imagine, because even if a gate length of 1 nm becomes plausible, the
extrapolation of the upper curve of \fig{f:rolloff}(a) implies an extremely
large gate voltage swing of around 5 V.

Decreasing the channel thickness $t$ would have a desirable effect on
the electrostatics of both thick- and thin-oxide devices. However, it
seems that the overall effect of reducing $t$ below 2 nm would be
deteriorating. First, in layers of such small thickness, the mobility
of electrons is expected to decrease sharply with decreasing
$t$\cite{Taur 97,Vasileska 97,Gamiz 98}. In fact, recent 
simulations\cite{Gamiz 98} predicted the electron mobility $\mu$ in SOI
MOSFET's to be around 350 cm$^2$/Vs at an effective electric field of
$6\times 10^5$ V/cm. 
This electric field corresponds to a confinement
potential with a first quantized level of width $t'=2$ nm. $\mu$ = 350
cm$^2$/Vs implies a scattering length $l \approx 20$ nm, which is
consistent with our ballistic model at $L < 15$ nm (it is also
consistent with the 
measurements of Ref.~\onlinecite{Huang 99} which find mobilities of the order
of 200 cm$^2$/Vs in MOSFET's with $t' \approx 1.5$ nm). However, at $t
\approx t' \ll 2$ nm, mobility is expected to becomes much smaller
than these values, implying a scattering length smaller than $L$,
which is inconsistent with our model, and which would deteriorate the
device due to strong backscattering.

It is worth emphasizing that for short devices ($L \approx 8$ nm),
tunneling current is large, and in fact may dominate over the thermal
current [\fig{f:separate}(b)]. In this sense, one can classify the short-channel devices
studied here as ``tunneling transistors''. The tunneling effect indeed
changes the overall shape of the current characteristics [\eg, the
subthreshold curve is no longer exponential, see
\fig{f:subthreshthick}(b)], but even in the strong-tunneling regime the
transistor is still responsive to gate voltage, enough to allow
practical current-control.

One important drawback of the devices studied here is the small (or
even negative)
threshold voltage $V_T$  (see
Figs.~\ref{f:subthreshthick},\ref{f:subthreshthin}). The main cause of
this effect (in addition to
the regular short-channel effects\cite{Taur 97,Fjeldly 93} which reduce
$V_T$ due to two-dimensional charge redistribution in the gate) is the
undoped channel. This leads to an 
accumulation of electrons in the channel starting at small negative
gate voltage (in contrast to regular n-channel MOSFET's with
p-type substrate in which electron accumulation in the channel is
possible only after 
the substantial depletion of holes by positive gate voltage.)

Two different approaches may be utilized to solve this problem (which
is of importance mainly to logic applications). One approach is to
allow a finite number of acceptor dopants in the channel. This would
have an effect similar to the p-substrate in regular MOSFET's, since
the gate voltage would first have to deplete the access holes before
allowing for accumulation of electrons. A crude estimate of this
effect can be obtained by using the planar capacitance model by which the
change in $V_T$ is given by
\[
\Delta V_T = {4 \pi q N_a t_{ox} \over \epsilon_{ox}},
\]
with $N_a$ the sheet density of acceptors, $\epsilon_{ox}$ the
dielectric constant of SiO$_{2}$, and $4 \pi t_{ox} / \epsilon_{ox}$
the gate capacitance (this approximation neglects any short-channel
effects). In order to achieve $\Delta V_T = 0.4 V$ (which would give
according to Fig.~\ref{f:subthreshthin} $V_T \approx 0.1 V$, which is
sufficiently large because of the small source-drain voltages in use),
$N_a$ should be approximately $6 \times 10^{12}$ cm$^{-2}$, which
implies $l \approx 4$ nm. Such channel doping is unacceptable since
the relation $l \sim L$ implies strong fluctuations in device
performance due to dopant fluctuations\cite{Bowen 99}.  Thus, it seems
that a more realistic approach to manipulating $V_T$ would be to use a
specific metal with necessary workfunction as a gate material.

The practical implementation of the remarkable MOSFET scaling
opportunities presented here requires several technological problems to be
solved. First of all, the fabrication of dual-gate transistors
requires rather advanced techniques - see,
{\it e.g.} Ref. \cite{Tiwari}.  Second, the gate voltage threshold $V_T$ of
nanoscale transistors is rather sensitive to nanometer fluctuations of
the channel length - see \fig{f:rolloff}(b).  Notice, however, that the {\it
relative} sensitivity of $V_T$ [which may be adequately characterized
by the log-log plot slope $(L/V_T) \times dV_T/dL$] {\it decreases} at
small $L$. This fact gives hope that with appropriate transistor
geometry (for example, vertical structures where $L$ is defined by
layer thickness rather than by patterning - see, {\it e.g.}, Ref.
\cite{Nakazato 99} ) the channel length fluctuations will eventually
be made small enough for appreciable VLSI circuit yields.

\section{Acknowledgments}

Helpful discussions with D.J. Frank, C. Hu, S. Laux, M. Fischetti, J.
Palmer, Y. Taur,
S. Tiwari, H.-S.P. Wong, R. Zhibin, and especially M. Lundstrom and P. M. Solomon
are gratefully 
acknowledged. This work was supported in part by the AME program of
DARPA via ONR.

\end{multicols}


\begin{thebibliography}{99}
\bibitem{neglecttunnel} The exponentially small density of tunneling
electrons below the potential barrier is neglected in our actual
solution of the Poisson equation.
\end{thebibliography}
\end{document}